\documentclass[12pt]{article} 

\usepackage{scicite}

\usepackage{times}

\usepackage{amsmath}
\usepackage{amsfonts}
\usepackage{amssymb}
\usepackage{graphicx}
\usepackage{subcaption}
\usepackage{placeins}
\usepackage{booktabs}
\usepackage{booktabs}

\newenvironment{sciabstract}{%
\begin{quote} \bf}
{\end{quote}}

\usepackage[labelfont=bf,figurename=Fig.,labelsep=period,font=footnotesize]{caption}

\title{How neighbourhood ideology shapes misinformation belief in densely tied social networks}

\author
{Soroush Karimi,${}^{1\ast}$ Marcos Oliveira,${}^{1,2}$ and Diogo Pacheco${}^{1}$\\
\\
\normalsize{${}^{1}$Department of Computer Science, University of Exeter, UK}\\
\normalsize{${}^{2}$Vrije Universiteit Amsterdam, Amsterdam, Netherlands}\\
\normalsize{$^\ast$\href{sk931@exeter.ac.uk}{sk931@exeter.ac.uk}}
}

\date{}

\usepackage{changepage}

\usepackage{xcolor}
\usepackage{soul}
\definecolor{reddish}{HTML}{FBB4AE}
\definecolor{blueish}{HTML}{B3CDE3}
\definecolor{magentish}{HTML}{FF00AA}
\definecolor{greenish}{HTML}{a1d99b}

\usepackage{hyperref}

\usepackage{setspace}

\usepackage{lipsum}

\begin{document} 




\maketitle 


\begin{sciabstract}
Abstract.
\textnormal{%
With the rapid spread of news on social media, understanding the propagation of misinformation is becoming increasingly important. One factor that affects individuals' vulnerability to false information is their ideological predisposition. Despite the large number of agent-based models that focus on social influence as a driver of the spread of false claims, they often fail to explicitly integrate personal ideological biases into belief formation. In this work, we explore how misinformation spreads through the interaction between individuals' ideological biases and social influence. Our model accounts for both the strength of individuals' ideological biases and the extent to which a false claim aligns with their ideology. Social influence modifies the effects of ideological intensity and false claim alignment through network interactions. Notably, the influence of neighbours' ideological intensity on belief is strongly affected by how well those neighbours are connected to one another. These results highlight the importance of considering both network structure and personal ideological biases when modelling misinformation propagation.
}
\end{sciabstract}

\baselineskip1.55em 

\section*{Introduction}

The mechanism of belief formation and information sharing is a social activity ~\cite{chung2016examining}, while personal biases play a key role in the way we process information~\cite{van2018partisan}. The impact of ideology and social influence has traditionally been decoupled in simulation models, and emerging unified frameworks have begun to integrate individual bias with network exposure~\cite{sikder2020minimalistic}. However, current models that combine the effects of social influence and personal bias lack an explicit mechanism that includes a range of bias toward an ideology and belief formation toward false claims from opposite to aligned.

Social media platforms provide an environment where individuals with varying levels of ideological intensity encounter false claims that differ in their ideological alignment. They now serve as news sources~\cite{fletcher2025link} and may increase the spread of unreliable information~\cite{allcott2017social} and toxic content~\cite{chetty2018hate}. Individuals receive content primarily from their social connections, and such exposure has a considerable impact on belief formation~\cite{pennycook2021psychology,puthineedi2026ball}. However, network and social influence alone do not explain all individual differences in what individuals believe and share~\cite{buchanan2020people}. 

One factor contributing to differences in belief and sharing is partisan or ideological bias (i.e., motivated reasoning)~\cite{van2022misinformation}. Experimental evidence shows that people systematically trust information more when it aligns with their political identity~\cite{thaler2024fake}. Ideological thinking is not restricted to political attitudes but is a broader domain-general cognitive style that varies along a spectrum from moderation to extremism~\cite{zmigrod2022psychology}. Together, these findings suggest that models of misinformation diffusion should account for variation in individuals' ideological intensity.

Simulation models, such as agent-based and epidemic contagion models, have been used to analyse the spread of misinformation~\cite{raponi2022fake}, as they can capture social influence through feedback loops that reinforce acceptance of a false claim~\cite{van2022misinformation}. More recently, frameworks have been developed that incorporate ideology into these contagion dynamics to better capture belief formation. For instance, Stein et al.~\cite{stein2023network} demonstrated that, in ideologically segregated networks, the prevalence of misinformation is higher in both experimental results and simulations, underscoring the joint importance of individual ideology and network topology. To capture the interaction aspect, Brooks et al.~\cite{brooks2020model} examined how media accounts with fixed ideological positions shape the spread of online content, modelling ordinary users who adopt ideologies and share content strictly within a designated ``ideology band''. Similarly, tracking multi-topic dynamics, Rodriguez et al.~\cite{rodriguez2016collective} combined psychological and social factors into a single framework to observe how public opinion transitions across multiple topics change. Their model explains how macro-level social phenomena, such as societal upheavals and the entrenchment of fringe groups, emerge from the interaction of cognitive and social forces. Moreover, in confirmation-bias models~\cite{del2017modeling}, an agent’s opinion and the information they encounter are treated as identical, thereby explaining how multiple beliefs can coexist. More explicit models have begun to develop to capture how bias shapes single-topic diffusion in the presence of cognitive forces. For example, Sikder et al.~\cite{sikder2020minimalistic} introduced a social learning model in which a fixed parameter determines an agent's probability of rejecting incongruent information. However, these frameworks fail to account for the interplay among social influence, individual ideological bias, and the ideological alignment of a specific false claim. Moreover, existing models lack a mechanism for representing how individuals with different levels of ideological intensity respond to false claims based on the claims' ideological alignment with their worldview.

In this work, we address the gap by explicitly focusing on belief formation about a false claim, integrating individual biases and social impact to determine whether individuals believe it. Unlike confirmation-bias models, by distinguishing between personal biases and the claim's ideological alignment, we can capture how behaviours differ based on the claim's ideological leaning. The distinction allows the model to describe the spread of misinformation. We show how network structure shapes belief dynamics and how the distribution of extremists alters social impact. We find that the effect of social influence is most pronounced in networks with dense communities. It implies that, beyond personal biases and the people we interact with, the structure of the community we belong to may increase our vulnerability to believing and sharing misinformation.

\section*{The Ideological Misinformation Contagion Model}
We introduce a contagion model that integrates individual ideological predisposition with social influence as joint drivers of misinformation propagation. Susceptibility to misinformation is affected by ideological congruence, as individuals tend to be more receptive to information that aligns with their existing beliefs and more sceptical of counter-attitudinal content~\cite{stein2024partisan,facciani2024personal}. Psychological research indicates that the strength of ideological commitment differs among individuals~\cite{zmigrod2022psychology}. We capture this variation by assigning each individual an ideological intensity score, which indicates the extent to which their prior ideological beliefs shape their reaction to information.

Belief formation can also be reinforced through social interactions, since exposure to like-minded peers can strengthen existing biases and promote the diffusion of misinformation~\cite{stein2024partisan}. We use the SBFC model~\cite{tambuscio2015fact} as the basis for our framework because it represents misinformation propagation through social interactions. We extend this framework by distinguishing an individual's ideological intensity from the ideological alignment of the false claim, and by incorporating their interaction into the transition dynamics. This allows the same individual to respond differently to different false claims according to how strongly each claim aligns with their ideology, while social interactions determine how these individual-level predispositions propagate through the network.

In broad terms, the model describes the propagation and rejection of a single false claim over a social network (hereafter, we refer to a `false claim' simply as a `claim'). Individuals can transition between susceptible, believer, and fact-checker states as a result of local social influence and the ideological relationship between the individual and the claim. This formulation allows the same individual to respond differently to different claims depending on their ideological alignment, while also allowing neighbouring agents to influence the individual's belief dynamics. For more details on the model, agent states and transitioning among them, see the Method Section.

\section*{Results}

The proposed model suggests that individual ideology and social influence jointly influence misinformation propagation. The simulation results further show that, in addition to an individual's ideological intensity, the influence of neighbours becomes increasingly important in highly clustered networks. We begin by examining the impact of ideological alignment of a claim on the overall percentage of believers. Then, we investigate belief formation at the node level to show how individuals respond differently to claims with different levels of ideological alignment. Lastly, we quantify the relative contributions of personal ideology and neighbours' ideological intensity across different network structures.

\subsection*{Greater Ideological Alignment Leads to More Believers}

To examine how the ideological alignment of a claim influences its spread, we compare how likely individuals are to believe it. We evaluate the percentage of believers across a continuous range of claim alignment, from completely opposite (\(A_{\text{factor}} = -1\)) to completely aligned (\(A_{\text{factor}} = 1\)), using an Erdős--Rényi random network (see Methods).

\begin{figure}[ht!]
    \centering
    \includegraphics[width=0.6\linewidth]{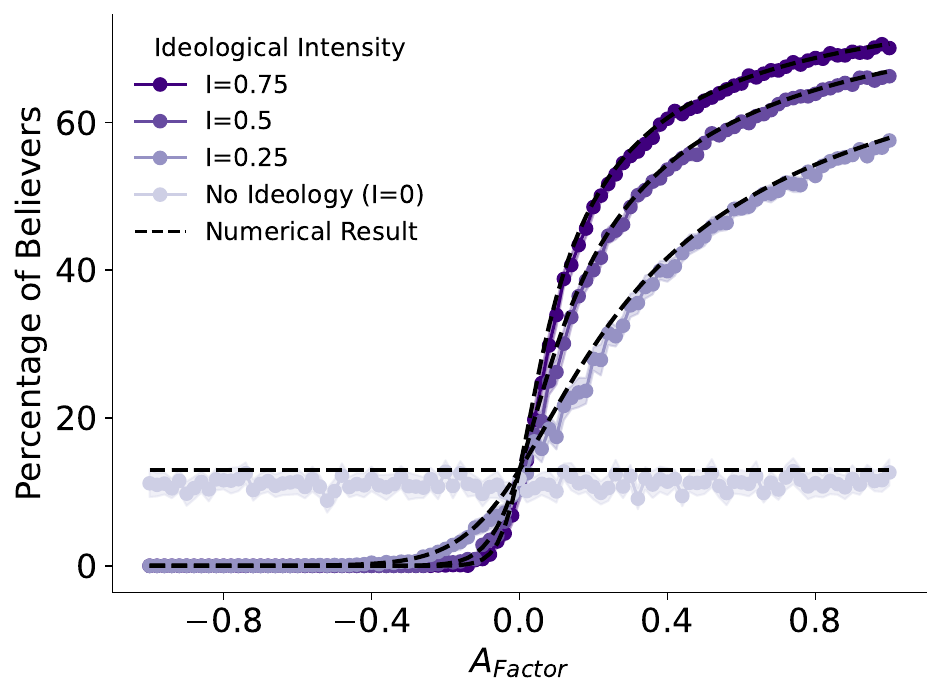}
    \caption{When the ideological intensity level is greater than zero, the percentage of believers increases with claim alignment. The ideological alignment factor indicates how closely the claim aligns with a specific ideology (e.g., right-wing). The simulation results align well with the numerical solution. }
    \label{fig:simulation_numerical}
\end{figure}

We find that when the claim is neutral with respect to individuals' ideology (\(A_{\text{factor}} = 0\)), individuals' ideological intensity does not affect the percentage of believers. When the claim is ideologically aligned with the individual (\(A_{\text{factor}} > 0\)), individuals are more likely to believe a claim, and the percentage of believers increases with ideological intensity (see Fig.~\ref{fig:simulation_numerical}). By contrast, for opposing claims (\(A_{\text{factor}} < 0\)), the percentage of believers decreases with ideological intensity. We complement the simulation results with a numerical solution of the Markov chain and an analytical treatment that formally characterises the steady-state belief dynamics. Both approaches closely match the simulation results.

\begin{figure}[t]
    \centering
    \includegraphics[width=0.99\linewidth]{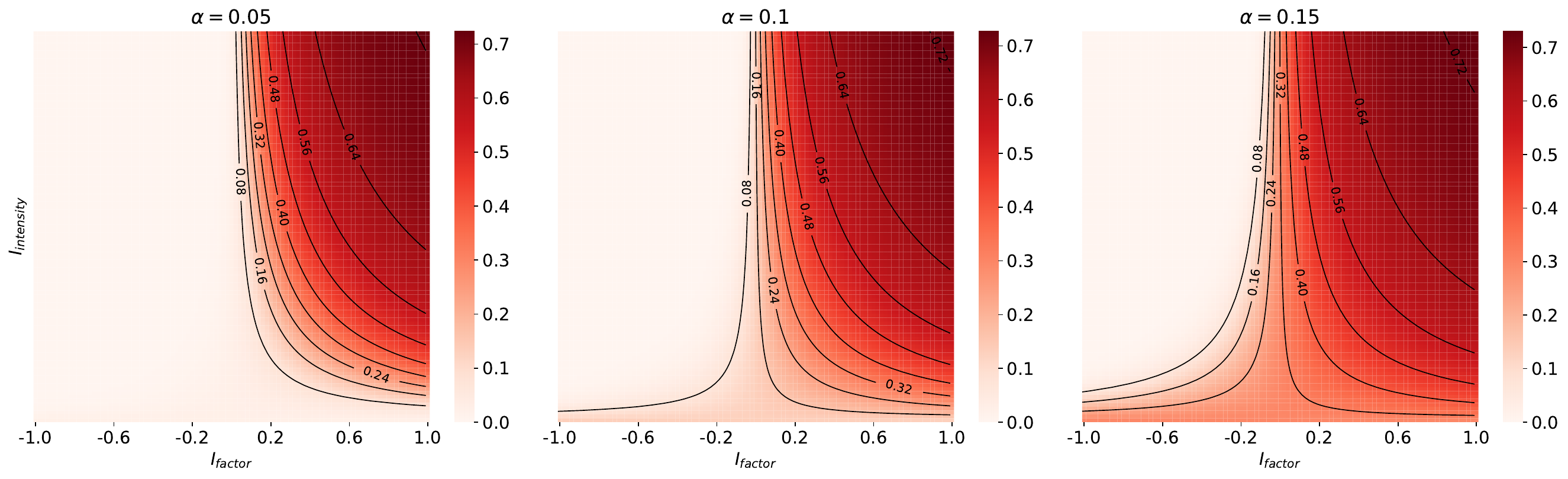}
    \caption{The fraction of believers as a function of ideological alignment factor (x-axis) and ideological intensity (y-axis). }
    \label{fig:Ideology_effect_heatmap}
\end{figure}

To understand how individual ideological strength and claim alignment influence belief, we examine how these effects vary with individuals' baseline vulnerability to claims, which we refer to as gullibility. In the low-gullibility case ($\alpha = 0.05$), few moderate individuals believe the claim, and greater ideological intensity is required for them to become believers. At higher gullibility levels, a fraction of moderate individuals become believers even at low levels of ideological intensity, reflecting their greater baseline gullibility (Fig.~\ref{fig:Ideology_effect_heatmap}). By comparing the panels at low ideological intensities, we observe that increasing $\alpha$ increases the probability that moderate individuals become believers.

\subsection*{Individual Belief Responses Vary with Claim Ideological Alignment}

To investigate how ideology and social influence affect individuals' beliefs about a claim, we examine each node separately. We conduct a node-level analysis by measuring the proportion of simulation runs that terminate in the believer state. We define this measurement as the belief probability and capture individuals' behaviour towards the claim. In this context, individuals are classified as either moderates, lacking strong ideological biases, or extremists, who possess high ideological intensity. Most extremists are grouped within a single cluster.

\begin{figure}[t]
    \centering
    \includegraphics[width=0.95\linewidth]{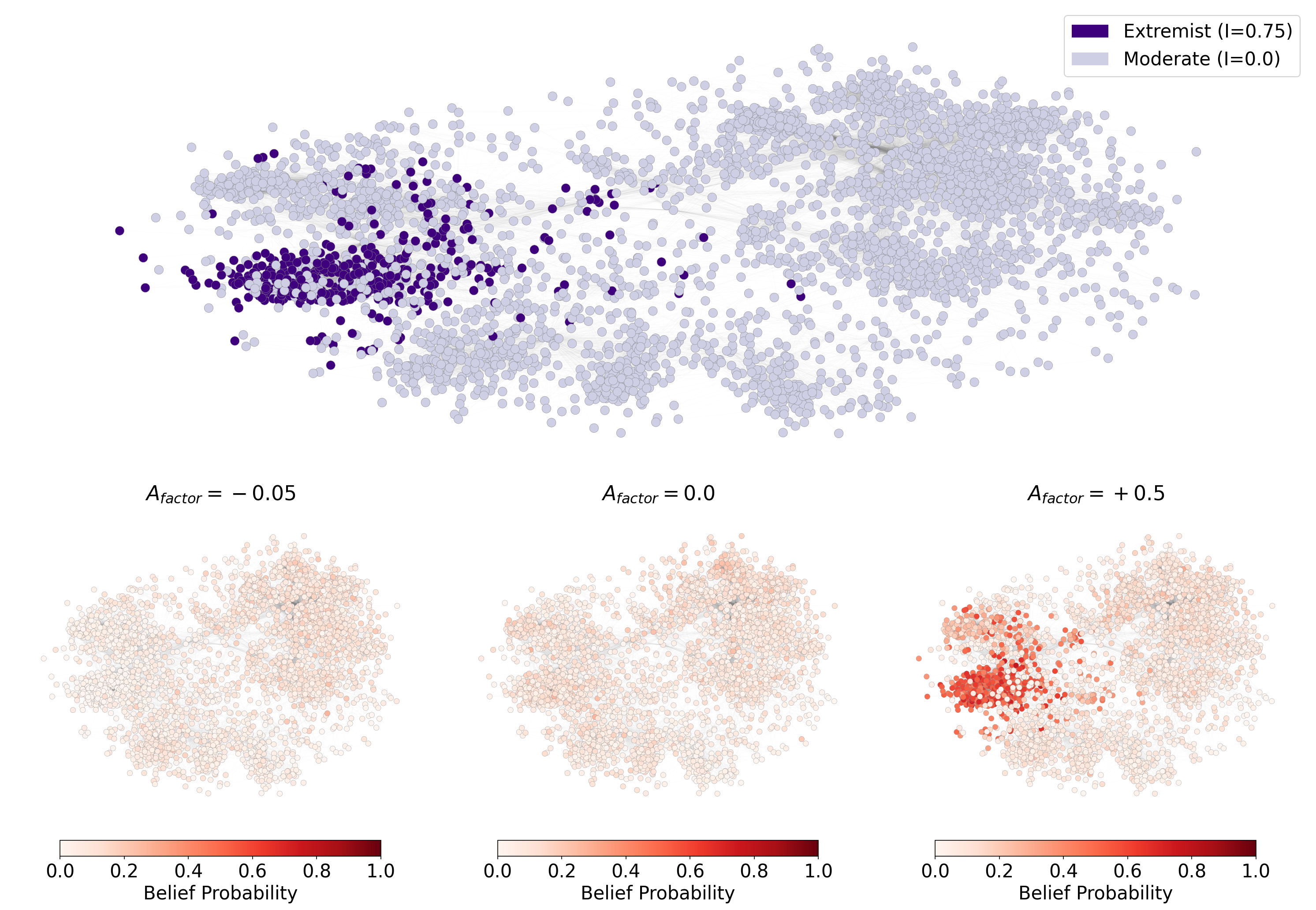}
    \caption{Different claims, different acceptance rates: At the top, we see the distribution of extremists, which are mostly clustered. In the second row, there are belief percentages for each node for cases where the claim is against, neutral toward, or aligned with the node's ideology. }
    \label{fig:network}
\end{figure}

We compare the belief probability for three types of claims: opposing, neutral, and aligned with the nodes' ideology. We find that when the claim opposes the nodes' ideology, extremists predominantly become fact-checkers, resulting in low belief probabilities. For neutral claims, ideological intensity has little effect, and moderates and extremists behave similarly. In contrast, when the claim aligns with the nodes' ideology, extremists are more likely to become believers. Additionally, as moderates are connected to extremist neighbours, their belief probabilities also increase (see Fig.~\ref{fig:network}). In the next subsection, we investigate how this neighbourhood effect varies across different strategies for distributing extremists within the networks.

\subsection*{Network Clustering Amplifies the Influence of Neighbourhood Ideology}

To understand how network structure affects social influence, we examine belief dynamics across four different network topologies and three strategies for placing extremists. Since an individual's state transitions are shaped by their neighbours, the positioning of extremists alters the ideological makeup of local communities. For each scenario, we define the belief probability as the proportion of simulation runs in which a node ends up in the believer state. We then explore how this probability is influenced by the node's own ideological strength and the average ideological intensity of its neighbours.

We sample ideological intensities uniformly (\(I_{\text{intensity}} \in [0,1]\)) and classify the top 30 per cent as extremists. Extremists are distributed using three topological strategies: uniformly at random, preferentially on high-degree hubs, and via breadth-first search (BFS) to form tight local clusters (see Methods). We evaluate belief formation under a claim that is strongly aligned with individuals' ideology (\(A_{\text{factor}}=0.8\)).

\begin{figure}[t!]
    \centering
    \includegraphics[width=1\linewidth]{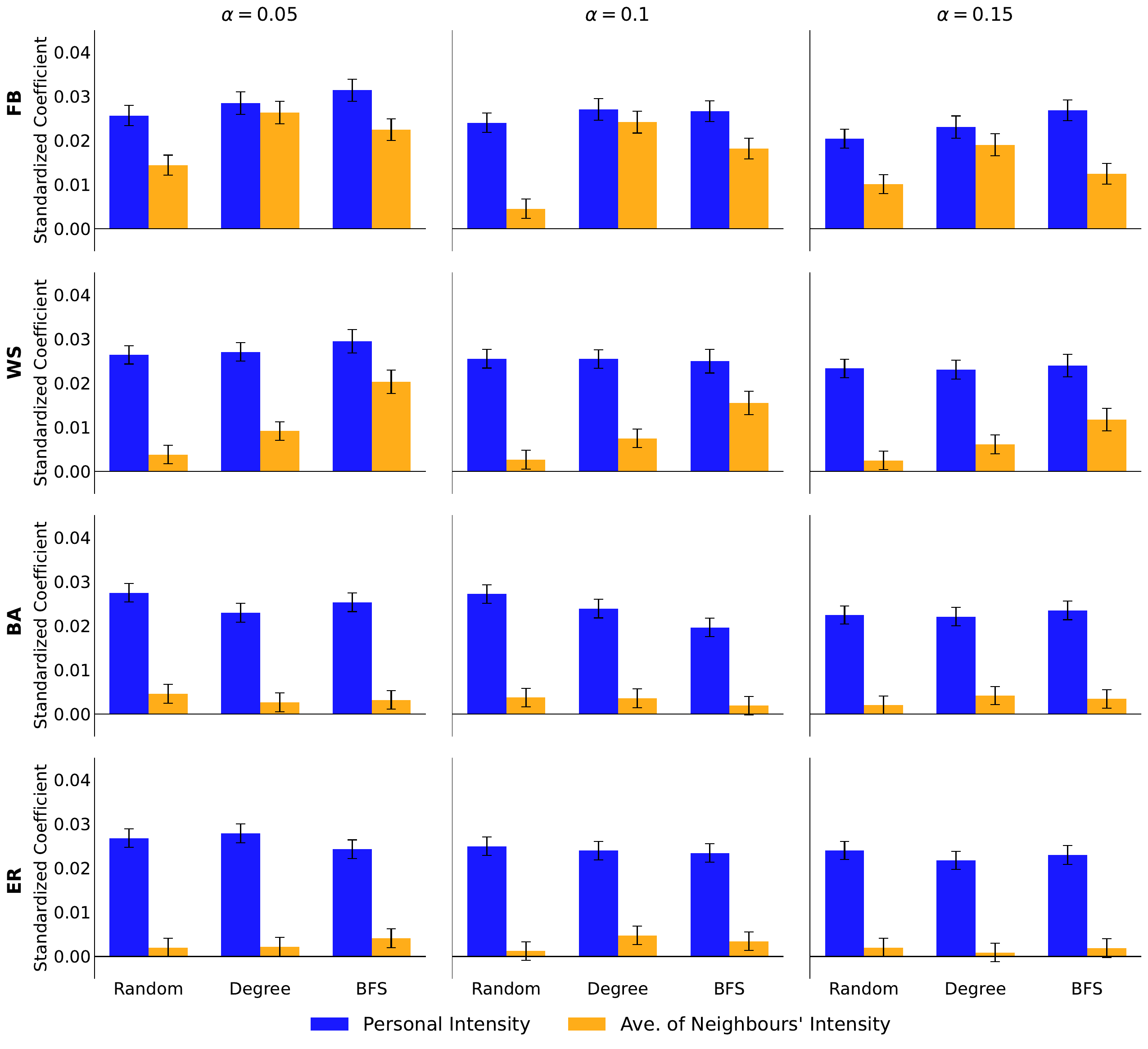}
    \caption{The impact of personal and neighbourhood ideology on belief probability using standardised regression coefficients. Personal intensity has a robust impact across different network structures and decreases with increasing general gullibility. The role of neighbours' intensity is greater in network structures with high clustering coefficients (e.g., Facebook and Watts-Strogatz). }
    \label{fig:personal_and_neighbours_impact}
\end{figure}

We observe that the influence of neighbours becomes much more significant in networks where neighbouring nodes are highly interconnected, as indicated by higher clustering coefficients (Table~\ref{tab:network_stats}). Networks like Facebook (average clustering coefficient of 0.606) exemplify strong clustering, which is closely matched by the Watts--Strogatz (WS) model (0.537), whereas models like Barabási--Albert (BA) and Erdős--Rényi (ER) have much lower values (0.037 and 0.011, respectively). Although Facebook and WS differ in degree distribution and community structure, both show a noticeable increase in the standardised regression coefficient for neighbours' ideological strength (Fig.~\ref{fig:personal_and_neighbours_impact}).
\begin{table}[htbp]
\centering
\begin{tabular}{lrrrr}
\toprule
\textbf{Structure} & \textbf{$N$} & \textbf{Edges} & \textbf{Avg. Degree ($\langle k \rangle$)} & \textbf{Avg. Clustering Coeff.} \\
\midrule
FB (empirical) & 4039 & 88{,}234 & 43.7 & 0.606 \\
WS             & 4039 & 88{,}858 & 44.0 & 0.537 \\
BA             & 4039 & 88{,}374 & 43.8 & 0.037 \\
ER             & 4039 & 88{,}234 & 43.7 & 0.011 \\
\bottomrule
\end{tabular}
\caption{Node count, edge count, average degree, and average local clustering
coefficient for each network structure used in the node-level comparison.
WS, BA, and ER were generated to match FB's node count and mean degree
(see Networks section).}
\label{tab:network_stats}
\end{table}

We find that in highly clustered networks, local peer pressure can match personal ideological bias in driving misinformation adoption. While personal ideological intensity is explicitly encoded in the transition probabilities, neighbourhood ideological intensity is not; this amplification emerges naturally from repeated local interactions constrained by network topology. Whereas network density has previously been shown to increase belief via baseline gullibility (\(\alpha\))~\cite{karimi2025modelling}, our result reveals how dense local communities specifically amplify the influence of neighbouring agents' ideologies. Finally, as general gullibility (\(\alpha\)) increases, individuals are generally more likely to become believers, while the relative effects of personal ideology and neighbours' ideological intensity are reduced.

\section*{Conclusion}

In this work, we introduce a contagion model that integrates individuals' ideological intensity with a claim's ideological alignment. While previous frameworks failed to account for this interaction alongside social influence, we show that dense network clustering endogenously amplifies the influence of neighbouring ideology, driving misinformation adoption beyond individual predispositions alone. Including individual biases provides a more plausible view of how misinformation spreads across the network. The model was implemented in various network structures to examine the impact of ideology alignment and individual biases. The simulations show that when individuals are unbiased, they behave similarly towards the claim, regardless of how aligned it is with their ideology. Claims that go against extremist values are debunked, and claims aligned with their ideology are believed. Furthermore, simulation results indicate that even small populations of highly ideological agents can significantly increase the adoption of misinformation among neighbouring moderates, especially within clustered network structures. This amplification is not directly encoded in the transition rules; rather, it arises from the interaction between ideological alignment and network topology. Here, we considered a single dimension of ideology and its alignment; in future work, there is room to study cases where individuals have multi-dimensional ideologies and claims with different alignment factors for each dimension. 

\section*{Methods}
\subsection*{Model}
In our model, agent occupies one of three discrete states regarding the misinformation item: a susceptible agent ($S$) is not aware of the item, a believer ($B$) thinks it is true, and a fact-checker ($F$) is aware that it is a claim and debunks it. 
To build the connection structure of the agents, we use a static, undirected social network $G=(V, E)$, where the nodes $i \in V$ represent individuals and the edges $E$ represent social ties. We define $n_i^{B}(t)$ and $n_i^{F}(t)$  as the numbers of neighbours of individual $i$ who are believers and fact-checkers, respectively, at time $t$.

\begin{figure}[t!]
    \centering
    \includegraphics[width=0.8\linewidth]{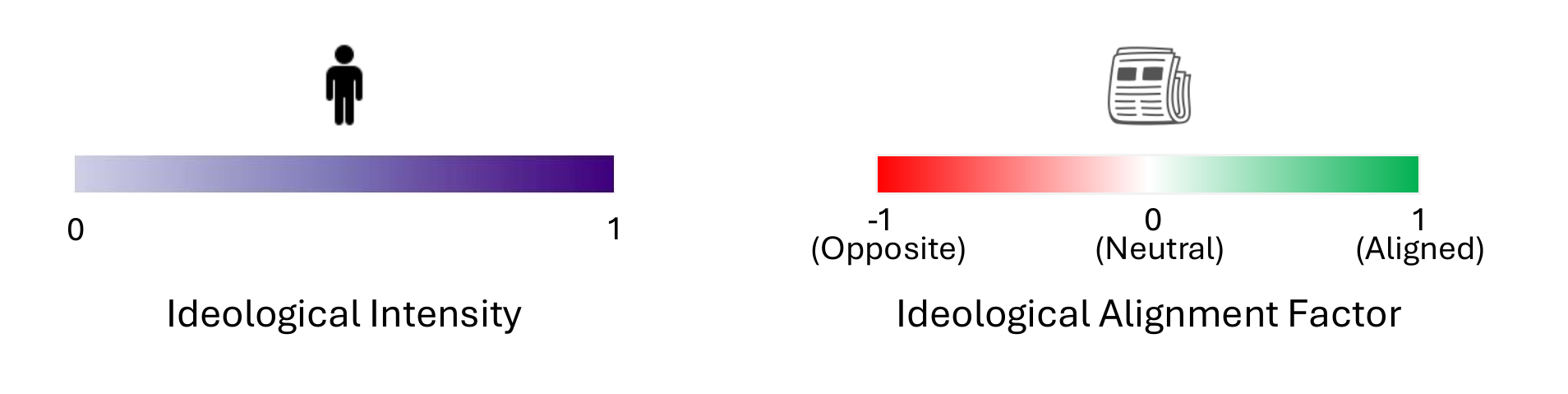}
    \caption{
    Ideological intensity is an individual attribute that quantifies the strength of an individual's ideological commitment ($0$-$1$). Meanwhile, the ideological alignment factor ($A_{\text{\text{factor}}}$) is an attribute of a claim that indicates its position on the ideological spectrum, ranging from $-1$ (completely opposite) to $1$ (fully aligned), with $0$ representing neutral content toward that specific ideology.
    }
    \label{fig:Ideology_term}
\end{figure}

We include ideological mechanisms whereby individuals are more likely to accept misinformation aligned with their worldview and reject content that conflicts with it. For each node $i$, the ideological intensity $I_i \in [0,1]$ specifies the strength of $i$'s commitment. The ideological alignment of the claim is represented by $A_{\text{factor}} \in [-1,1]$. A value of $A_{\text{factor}}=-1$ represents content that is completely opposed to the relevant ideology, $A_{\text{factor}}=0$ represents ideologically neutral content, and $A_{\text{factor}}=1$ denotes content that is fully aligned with it. Thus, $I_i$ is an individual-level attribute, while $A_{\text{factor}}$ is a property of the misinformation (Fig.\ref{fig:Ideology_term}). By this distinction, the same individual responds differently to claims with different ideological alignments.

A positive value of $A_{\text{factor}} I_i$ indicates ideological alignment, a negative value indicates ideological mismatch, and a value close to zero indicates that ideological considerations have little influence on the individual's response. In the multidimensional case, this scalar product can be extended to a dot product between an individual's ideological-intensity vector and the claim's alignment vector.
In addition to ideological motivation for either believing or debunking a claim, there is a certain level of vulnerability to believing the claim, which is determined by $\alpha\in[0,1]$. 
We define
\[
L_i=\frac{\alpha + A_{\text{factor}}I_i}{2}
\]
which combines the baseline vulnerability with ideological motivation. The spreading rate is determined by $\beta\in[0,1]$. The probability that a susceptible individual becomes a believer is equal to:
\begin{equation}
    f_i(t) = \beta \frac{n_i^B(t) (1 + L_i)}{n_i^B(t) (1 + L_i) + n_i^F(t) (1 - L_i)},\label{equation:f}
\end{equation}

The corresponding probability of transitioning from susceptible to fact-checker is:

\begin{equation}
    g_i(t) = \beta \frac{n_i^F(t) (1 - L_i)}{n_i^B(t) (1 + L_i) + n_i^F(t) (1 - L_i)}.\label{equation:g}
\end{equation}

When the claim is aligned with an individual's ideology ($A_{\text{factor}}>0$), greater ideological intensity increases their relative susceptibility to believer influence. Conversely, when opposed ($A_{\text{factor}}<0$), greater ideological intensity decreases $L_i$ and increases the relative weight placed on fact-checker influence. For ideologically neutral content ($A_{\text{factor}}=0$), there is no difference between nodes with different ideological intensities, which reduces the framework to the baseline SBFC misinformation model~\cite{tambuscio2015fact}.

\begin{figure}[h!]
    \centering
    \includegraphics[width=0.5\linewidth]{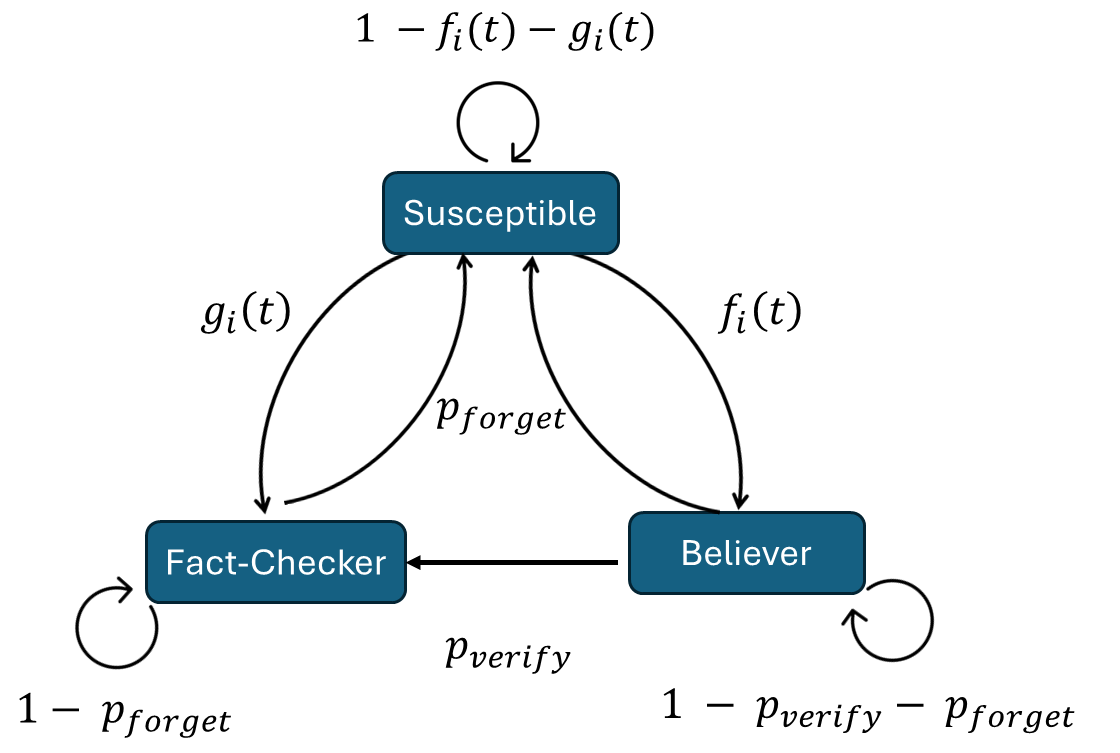}
    \caption{Transition diagram}
    \label{fig:transition-diagram}
\end{figure}%

In this model, individuals may change their state after initially adopting or rejecting the claim with fixed probabilities. A believer can verify the information and become a fact-checker with probability $p_{\mathrm{verify}}$, halting the spread of the claim and contributing to debunking.
Both believers and fact-checkers can also forget their current position towards the claim and return to the susceptible state with probability $p_{\mathrm{forget}}$ (Fig.\ref{fig:transition-diagram}). In this way, the model permits exposure to the claim and transitions between states.

\subsection*{Simulation}

We performed simulations on static, undirected networks across discrete time steps. Each node could occupy one of three states: susceptible ($S$), believer ($B$), or fact-checker ($F$). We initialised node states with 98\% susceptible, 1\% believer, and 1\% fact-checker.

At each discrete time step, transition probabilities were determined based on the transition rates shown in Fig.~\ref {fig:transition-diagram}. We set the model parameters to $\beta=0.5$, $\alpha=0.1$, $p_{verify}=0.01$ and $p_{forget}=0.1$. We selected the parameters in the range that avoids trivial absorbing states (e.g., complete elimination of believers) so that the influence of ideology and social interactions on belief formation can be analysed. Each simulation was run for 500 time steps, which exceeded the number of steps required to reach a steady state.

To investigate the influence of ideological alignment, the alignment factor was varied from $-1$ to $1$, while ideological intensity took the values $0$, $0.25$, $0.5$, and $0.75$. For each parameter combination, ran 50 trials. The reported percentage of believers is the mean, and the uncertainty is given as the 95\% confidence interval.

\subsection*{Markov Representation}
In addition to agent-based simulations, we provide complementary numerical and analytical solutions (see the appendix) to validate our simulation findings and formally characterise the steady-state belief dynamics.
Based on this model, in which the current state of each individual is determined solely by its previous state, we constructed a Markov chain in which the transition rates from the susceptible state to the believer and susceptible-to-fact-checker states are functions of the probabilities of being in different states.
Eq.~\eqref{eq:transition-matrix} shows the transition matrix and how the probability of being in each state updates.

\begin{equation}
    \label{eq:transition-matrix}
    \begin{bmatrix}
    P^S(t+1) \\
    P^F(t+1) \\
    P^B(t+1)
    \end{bmatrix}
    =
    \begin{bmatrix}
    1 - f_i(t) - g_i(t) & p_{\text{forget}} & p_{\text{forget}} \\
    g_i(t) & 1 - p_{\text{forget}} & p_{\text{verify}} \\
    f_i(t) & 0 & 1 - p_{\text{forget}} - p_{\text{verify}}
    \end{bmatrix}
    \begin{bmatrix}
    P^S(t) \\
    P^F(t) \\
    P^B(t)
    \end{bmatrix}
\end{equation}

\subsubsection*{Statistical Analysis}

At the node level, we calculated the belief probability as the proportion of simulation runs in which the node ended in the believer state. We used ordinary least squares (OLS) regression models to quantify the relative contributions of a node's ideological intensity and the mean ideological intensity of its neighbours to belief probability. To compare the regression coefficients directly, we standardised the predictor variables (z-scores) before fitting the model. We obtained confidence intervals for the regression coefficients from the fitted models. We repeated the procedure for different network structures and three extremist placement strategies at fixed $A_{\text{factor}}= 0.8$ and across three values of $\alpha$ (0.05,0.1,0.15).

\subsubsection*{Networks}

For the baseline simulations (Fig.~\ref{fig:simulation_numerical}), we generated Erdős-Rényi (ER) networks with $N = 1000$ nodes and edge probability $p = 0.04$ (average degree $\langle k \rangle \approx 40$). For the empirical network analysis, we used a network sample structure from Facebook~\cite{leskovec2012learning} comprising $N = 4039$ nodes with an average degree of $\langle k \rangle \approx 43.7$. For the comparative node-level analysis, we used the \texttt{NetworkX} Python package to generate synthetic network structures, including Erdős-Rényi (ER), Watts-Strogatz (WS), and Barabási-Albert (BA) models, preserving a closely matched number of nodes ($N = 4039$) and mean degree as the empirical Facebook network.

\subsection*{Ideological Intensity and Extremist Definitions}

\begin{figure}[h!]
    \centering
    \includegraphics[width=0.8\linewidth]{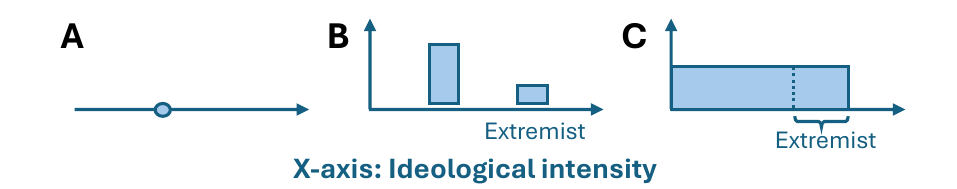}
    \caption{Ideological intensity choices. A) There is a single value for all the nodes. B) The binary case, when we have two intensity values, one for moderates and one for extremists. C) Ideological intensity forms a uniform distribution, in which the extremists are the nodes with the 30 per cent highest ideological intensity values.
}
    \label{fig:intensity_extermists}
\end{figure}

To analyse the impact of individual bias on misinformation dynamics, we consider three configurations of ideological intensity ($I_{intensity}$), as shown in Fig.~\ref{fig:intensity_extermists}:

\begin{enumerate}
    \setlength\itemsep{0pt}
    \setlength\parsep{0pt}
    \item \textbf{Homogeneous Baseline:} To examine the general effect of ideology in terms of believing a claim, we assume the same ideological intensity for all individuals (Fig.~\ref{fig:intensity_extermists}A). This reveals the relationship between a claim's ideological alignment ($A_{\text{factor}}$) and the global spread of misinformation.
    
    \item \textbf{Binary Case:} For demonstration purposes, we utilise a binary distribution in which nodes are strictly assigned to either a ''moderate'' or an ''extremist'' intensity value. This simplifies the observation of how extreme bias affects individual state transitions (Fig.~\ref{fig:intensity_extermists}B).
    
    \item \textbf{Continuous Heterogeneous Distribution:} For node-level analysis, ideological intensity values are selected from a uniform distribution in the interval $[0,1]$. We define \textbf{extremists} as the 30\% of the population with the highest intensity values (at or above the 70th percentile), while the remaining 70\% are moderates (Fig.~\ref{fig:intensity_extermists}C).  It is approximately equivalent to assigning ideological intensity values greater than 0.7 to extremists and the remaining values to moderates.
\end{enumerate}

In the heterogeneous scenario, a single set of values is sampled from this distribution for the entire population and ranked; the top 30\% form the extremist pool and the remainder form the moderate pool. The values within these pools are shuffled before being assigned to nodes. Thus, the extremists are always more intense than the moderates, but the exact value does not depend on the network position, unless a specific placement strategy (e.g. degree-based or BFS-based placement) is adopted.

\subsection*{Extremist Placement Strategies}
\begin{figure}[h!]
    \centering
    \includegraphics[width=1\linewidth]{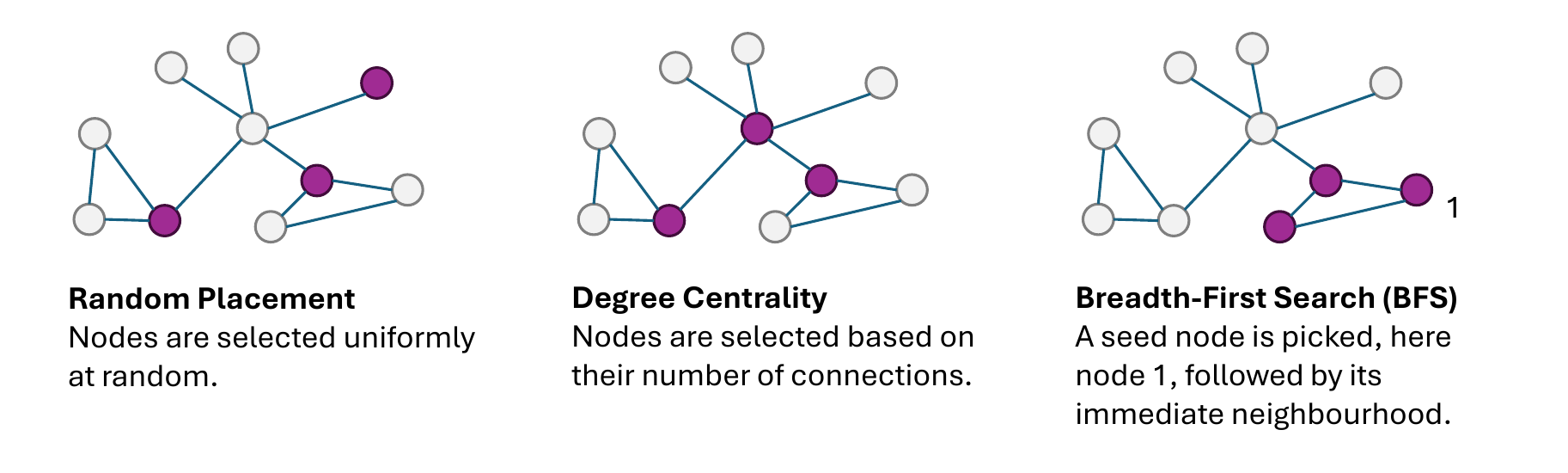}
    \caption{Extremist agents are seeded into the network according to three distinct topological strategies: Random Placement, Degree Centrality, and Localised (BFS) Search. 
}
    \label{fig:extremist_seed}
\end{figure}

Once the extremist nodes are defined, they are seeded into the network according to three distinct topological strategies (Fig.~\ref{fig:extremist_seed}).

\begin{enumerate}
    \setlength\itemsep{0pt}
    \setlength\parsep{0pt}
\item \textbf{Random Placement:} Extremist nodes are selected uniformly at random from the set of all nodes.
\item \textbf{Degree-Based Placement:} Nodes are ranked by degree centrality, and the $N_{ext}$ nodes with the most connections are designated as extremists. This strategy simulates a scenario in which ''hubs'' or influential individuals are the primary drivers of the ideology.

\item \textbf{Localized (BFS) Placement:} To simulate ideological clusters or ''echo chambers,'' we employ a Breadth-First Search (BFS) approach. A seed node is selected at random and assigned as an extremist, followed by its immediate neighbours (depth-1). If the required count of extremists is not met, a new random seed is chosen, and the process repeats until 30\% of the population is seeded.
\end{enumerate}

\bibliography{sciadvbib}

\begin{thebibliography}{10}
\expandafter\ifx\csname url\endcsname\relax
  \def\url#1{\texttt{#1}}\fi
\expandafter\ifx\csname urlprefix\endcsname\relax\def\urlprefix{URL }\fi
\providecommand{\bibinfo}[2]{#2}
\providecommand{\eprint}[2][]{\url{#2}}

\bibitem{chung2016examining}
\bibinfo{author}{Chung, N.}, \bibinfo{author}{Nam, K.} \& \bibinfo{author}{Koo, C.}
\newblock \bibinfo{title}{Examining information sharing in social networking communities: Applying theories of social capital and attachment}.
\newblock \emph{\bibinfo{journal}{Telematics and Informatics}} \textbf{\bibinfo{volume}{33}}, \bibinfo{pages}{77--91} (\bibinfo{year}{2016}).

\bibitem{van2018partisan}
\bibinfo{author}{Van~Bavel, J.~J.} \& \bibinfo{author}{Pereira, A.}
\newblock \bibinfo{title}{The partisan brain: An identity-based model of political belief}.
\newblock \emph{\bibinfo{journal}{Trends in cognitive sciences}} \textbf{\bibinfo{volume}{22}}, \bibinfo{pages}{213--224} (\bibinfo{year}{2018}).

\bibitem{sikder2020minimalistic}
\bibinfo{author}{Sikder, O.}, \bibinfo{author}{Smith, R.~E.}, \bibinfo{author}{Vivo, P.} \& \bibinfo{author}{Livan, G.}
\newblock \bibinfo{title}{A minimalistic model of bias, polarization and misinformation in social networks}.
\newblock \emph{\bibinfo{journal}{Scientific reports}} \textbf{\bibinfo{volume}{10}}, \bibinfo{pages}{5493} (\bibinfo{year}{2020}).

\bibitem{fletcher2025link}
\bibinfo{author}{Fletcher, R.} \emph{et~al.}
\newblock \bibinfo{title}{The link between changing news use and trust: longitudinal analysis of 46 countries}.
\newblock \emph{\bibinfo{journal}{Journal of Communication}} \textbf{\bibinfo{volume}{75}}, \bibinfo{pages}{1--15} (\bibinfo{year}{2025}).

\bibitem{allcott2017social}
\bibinfo{author}{Allcott, H.} \& \bibinfo{author}{Gentzkow, M.}
\newblock \bibinfo{title}{Social media and fake news in the 2016 election}.
\newblock \emph{\bibinfo{journal}{Journal of economic perspectives}} \textbf{\bibinfo{volume}{31}}, \bibinfo{pages}{211--236} (\bibinfo{year}{2017}).

\bibitem{chetty2018hate}
\bibinfo{author}{Chetty, N.} \& \bibinfo{author}{Alathur, S.}
\newblock \bibinfo{title}{Hate speech review in the context of online social networks}.
\newblock \emph{\bibinfo{journal}{Aggression and violent behavior}} \textbf{\bibinfo{volume}{40}}, \bibinfo{pages}{108--118} (\bibinfo{year}{2018}).

\bibitem{pennycook2021psychology}
\bibinfo{author}{Pennycook, G.} \& \bibinfo{author}{Rand, D.~G.}
\newblock \bibinfo{title}{The psychology of fake news}.
\newblock \emph{\bibinfo{journal}{Trends in cognitive sciences}} \textbf{\bibinfo{volume}{25}}, \bibinfo{pages}{388--402} (\bibinfo{year}{2021}).

\bibitem{puthineedi2026ball}
\bibinfo{author}{Puthineedi, V.} \& \bibinfo{author}{Jha, A.~K.}
\newblock \bibinfo{title}{Where the ball starts rolling? an empirical investigation into initial opinion formation on social media platforms}.
\newblock \emph{\bibinfo{journal}{Information Systems Research}}  (\bibinfo{year}{2026}).

\bibitem{buchanan2020people}
\bibinfo{author}{Buchanan, T.}
\newblock \bibinfo{title}{Why do people spread false information online? the effects of message and viewer characteristics on self-reported likelihood of sharing social media disinformation}.
\newblock \emph{\bibinfo{journal}{Plos one}} \textbf{\bibinfo{volume}{15}}, \bibinfo{pages}{e0239666} (\bibinfo{year}{2020}).

\bibitem{van2022misinformation}
\bibinfo{author}{Van Der~Linden, S.}
\newblock \bibinfo{title}{Misinformation: susceptibility, spread, and interventions to immunize the public}.
\newblock \emph{\bibinfo{journal}{Nature medicine}} \textbf{\bibinfo{volume}{28}}, \bibinfo{pages}{460--467} (\bibinfo{year}{2022}).

\bibitem{thaler2024fake}
\bibinfo{author}{Thaler, M.}
\newblock \bibinfo{title}{The fake news effect: Experimentally identifying motivated reasoning using trust in news}.
\newblock \emph{\bibinfo{journal}{American Economic Journal: Microeconomics}} \textbf{\bibinfo{volume}{16}}, \bibinfo{pages}{1--38} (\bibinfo{year}{2024}).

\bibitem{zmigrod2022psychology}
\bibinfo{author}{Zmigrod, L.}
\newblock \bibinfo{title}{A psychology of ideology: Unpacking the psychological structure of ideological thinking}.
\newblock \emph{\bibinfo{journal}{Perspectives on Psychological Science}} \textbf{\bibinfo{volume}{17}}, \bibinfo{pages}{1072--1092} (\bibinfo{year}{2022}).

\bibitem{raponi2022fake}
\bibinfo{author}{Raponi, S.}, \bibinfo{author}{Khalifa, Z.}, \bibinfo{author}{Oligeri, G.} \& \bibinfo{author}{Di~Pietro, R.}
\newblock \bibinfo{title}{Fake news propagation: a review of epidemic models, datasets, and insights}.
\newblock \emph{\bibinfo{journal}{ACM Transactions on the Web (TWEB)}} \textbf{\bibinfo{volume}{16}}, \bibinfo{pages}{1--34} (\bibinfo{year}{2022}).

\bibitem{stein2023network}
\bibinfo{author}{Stein, J.}, \bibinfo{author}{Keuschnigg, M.} \& \bibinfo{author}{van~de Rijt, A.}
\newblock \bibinfo{title}{Network segregation and the propagation of misinformation}.
\newblock \emph{\bibinfo{journal}{Scientific Reports}} \textbf{\bibinfo{volume}{13}}, \bibinfo{pages}{917} (\bibinfo{year}{2023}).

\bibitem{brooks2020model}
\bibinfo{author}{Brooks, H.~Z.} \& \bibinfo{author}{Porter, M.~A.}
\newblock \bibinfo{title}{A model for the influence of media on the ideology of content in online social networks}.
\newblock \emph{\bibinfo{journal}{Physical Review Research}} \textbf{\bibinfo{volume}{2}}, \bibinfo{pages}{023041} (\bibinfo{year}{2020}).

\bibitem{rodriguez2016collective}
\bibinfo{author}{Rodriguez, N.}, \bibinfo{author}{Bollen, J.} \& \bibinfo{author}{Ahn, Y.-Y.}
\newblock \bibinfo{title}{Collective dynamics of belief evolution under cognitive coherence and social conformity}.
\newblock \emph{\bibinfo{journal}{PLoS one}} \textbf{\bibinfo{volume}{11}}, \bibinfo{pages}{e0165910} (\bibinfo{year}{2016}).

\bibitem{del2017modeling}
\bibinfo{author}{Del~Vicario, M.}, \bibinfo{author}{Scala, A.}, \bibinfo{author}{Caldarelli, G.}, \bibinfo{author}{Stanley, H.~E.} \& \bibinfo{author}{Quattrociocchi, W.}
\newblock \bibinfo{title}{Modeling confirmation bias and polarization}.
\newblock \emph{\bibinfo{journal}{Scientific reports}} \textbf{\bibinfo{volume}{7}}, \bibinfo{pages}{40391} (\bibinfo{year}{2017}).

\bibitem{stein2024partisan}
\bibinfo{author}{Stein, J.}, \bibinfo{author}{Keuschnigg, M.} \& \bibinfo{author}{van~de Rijt, A.}
\newblock \bibinfo{title}{Partisan belief in new misinformation is resistant to accuracy incentives}.
\newblock \emph{\bibinfo{journal}{PNAS nexus}} \textbf{\bibinfo{volume}{3}}, \bibinfo{pages}{pgae506} (\bibinfo{year}{2024}).

\bibitem{facciani2024personal}
\bibinfo{author}{Facciani, M.} \& \bibinfo{author}{Steenbuch-Traberg, C.}
\newblock \bibinfo{title}{Personal network composition and cognitive reflection predict susceptibility to different types of misinformation}.
\newblock \emph{\bibinfo{journal}{Connections}} \textbf{\bibinfo{volume}{44}}, \bibinfo{pages}{165--180} (\bibinfo{year}{2024}).

\bibitem{tambuscio2015fact}
\bibinfo{author}{Tambuscio, M.}, \bibinfo{author}{Ruffo, G.}, \bibinfo{author}{Flammini, A.} \& \bibinfo{author}{Menczer, F.}
\newblock \bibinfo{title}{Fact-checking effect on viral hoaxes: A model of misinformation spread in social networks}.
\newblock In \emph{\bibinfo{booktitle}{Proceedings of the 24th international conference on World Wide Web}}, \bibinfo{pages}{977--982} (\bibinfo{year}{2015}).

\bibitem{karimi2025modelling}
\bibinfo{author}{Karimi, S.}, \bibinfo{author}{Oliveira, M.} \& \bibinfo{author}{Pacheco, D.}
\newblock \bibinfo{title}{{Modelling Misinformation Spread: The Role of Network Density in Diverse Social Structures}}.
\newblock In \bibinfo{editor}{Czupryna, M.}, \bibinfo{editor}{Kami{\'{n}}ski, B.} \& \bibinfo{editor}{Verhagen, H.} (eds.) \emph{\bibinfo{booktitle}{Advances in Social Simulation}}, Springer Proceedings in Complexity, \bibinfo{pages}{223--230} (\bibinfo{publisher}{Springer}, \bibinfo{address}{Cham}, \bibinfo{year}{2025}).

\bibitem{leskovec2012learning}
\bibinfo{author}{Leskovec, J.} \& \bibinfo{author}{Mcauley, J.}
\newblock \bibinfo{title}{Learning to discover social circles in ego networks}.
\newblock \emph{\bibinfo{journal}{Advances in neural information processing systems}} \textbf{\bibinfo{volume}{25}} (\bibinfo{year}{2012}).

\end{thebibliography}
\bibliographystyle{naturemag}




\section*{Appendix}

The main text focuses on agent-based simulations of the proposed model. In addition, when all individuals are assumed to have the same level of ideological intensity, the model admits a mean-field analytical treatment. 
Here we provide the analytical solution for the steady-state fraction of believers for the homogeneous-intensity case.

The corresponding state probabilities evolve according to
\begin{equation}
    p_i^B(t + 1) = f_i(t)P_i^S(t) + (1 - p_{forget} - p_{verify})P_i^B(t)
    \label{equation:pb}
\end{equation}

\begin{equation}
    p_i^F(t + 1) = g_i(t)P_i^S(t) + p_{verify}P_i^B(t) + (1 - p_{forget})P_i^F(t) 
    \label{equation:pf}
\end{equation}

\begin{equation}
    p_i^S(t + 1) = [1 - f_i(t) - g_i(t)]P_i^S(t) + p_{forget} P_i^B(t) + p_{forget} P_i^F(t). 
\end{equation}

At steady state, we have $P^X(t+1) = P^X(t) = P(X)$. In the mean-field approximation, the transition probabilities $f_i(t)$ and $g_i(t)$ are replaced by population-level expressions $f$ and $g$:
\begin{equation}
f = \beta \frac{P(B)(1+L)}{P(B)(1+L) + P(F)(1-L)}, \quad g = \beta \frac{P(F)(1-L)}{P(B)(1+L) + P(F)(1-L)}
\end{equation}
By rearranging equations \ref{equation:pb} and \ref{equation:pf} at the steady state we have:
\begin{align}
    f P(S) &= (p_{\mathrm{forget}} + p_{\mathrm{verify}}) P(B), \label{eq:balance_b} \\
    g P(S) + p_{\mathrm{verify}} P(B) &= p_{\mathrm{forget}} P(F). \label{eq:balance_f}
\end{align}

Furthermore, the sum of the probabilities must equal 1:
\begin{equation}
    P(S) + P(B) + P(F) = 1.
\end{equation}

Solving the equations for $P(B)$ results the steady-state fraction of believers:

\[
P(B) =
\begin{cases}
\dfrac{
\beta \left[ 2 L p_{\mathrm{forget}} - (1-L) p_{\mathrm{verify}} \right]
}{
2L(\beta + p_{\mathrm{forget}}) (p_{\text{forget}} + p_{\text{verify}})
},
&
\text{if } \beta \left[ 2L p_{\mathrm{forget}} - (1-L)  p_{\mathrm{verify}} \right] > 0
\\[12pt]
0, & \text{otherwise}
\end{cases}
\]

\begin{figure}[ht!]
    \centering
    \includegraphics[width=1\linewidth]{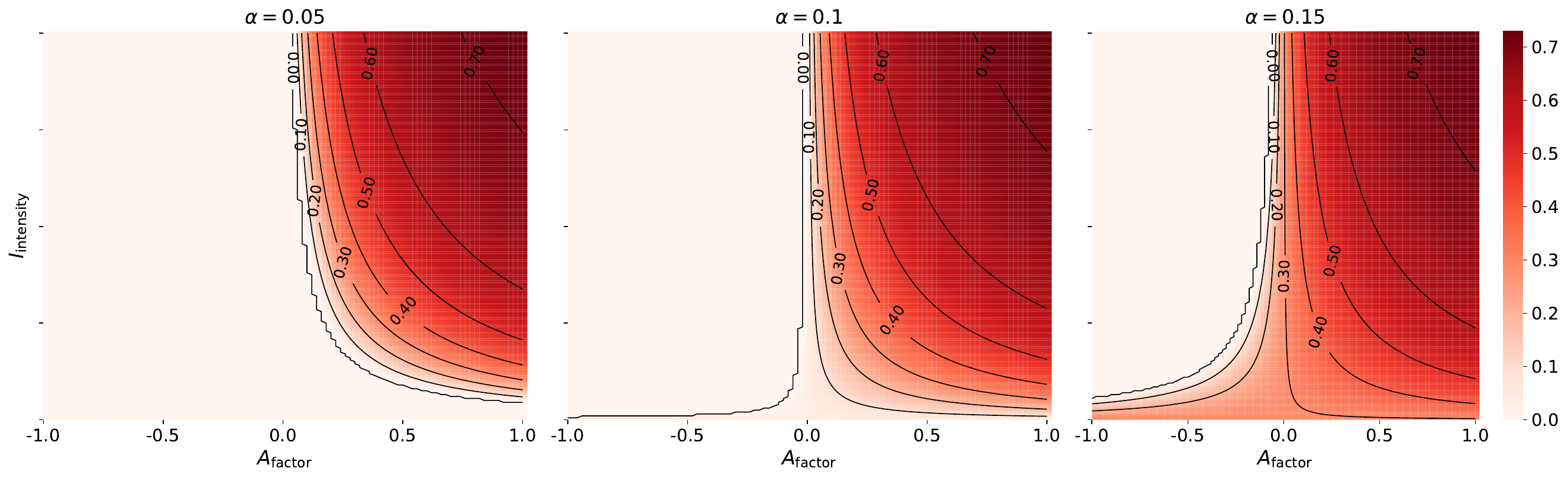}
    \caption{Fraction of believers as a function of ideological alignment factor and ideological intensity, obtained from the analytical solution.}
    \label{fig:analytical}
\end{figure}

The outcome of the analytical solution is shown in Fig.\ref{fig:analytical}.

\end{document}